\documentstyle[pra,aps,multicol,eqsecnum,epsfig]{revtex}

\def\be{\begin{eqnarray}}
\def\ee{\end{eqnarray}}

\begin{document}
\title{Quantum-state synthesis of multi-mode bosonic fields:
 Preparation of arbitrary states of 2-D vibrational motion of trapped ions}
\author{G. Drobn\'y$^{1}$, B. Hladk\'y$^{1,2}$ and V. Bu\v{z}ek$^{1,2}$}
\address{$^{1}$ Institute of Physics, Slovak Academy of Sciences,
		D\'ubravsk\'a cesta 9, 842 28 Bratislava, Slovakia\\
	 $^{2}$ Department of Optics, Comenius University,
		Mlynsk\'a dolina, 842 15 Bratislava, Slovakia
}

\date{April 22, 1998}
\maketitle

\begin{abstract}
We present a {\em universal} algorithm for an efficient {\em deterministic}
preparation of an {\em arbitrary} two--mode bosonic state.
In particular, we discuss in detail  preparation
of entangled states of a two-dimensional vibrational
motion of a trapped ion via a sequence of
laser stimulated Raman transitions.
Our formalism can be generalized for multi-mode bosonic
fields. We examine stability of our algorithm with respect to
a technical noise.

\end{abstract}

%\widetext
%\narrowtext
\begin{multicols}{2}

%%%%%%%%%%%%%%%%%%%%%%%%%%%%%%%%%%%%%%%%%%%%%%%%%%%%%%%%%%%%%%%%%%%%%%%%%%%%
\section{Introduction}
%%%%%%%%%%%%%%%%%%%%%%%%%%%%%%%%%%%%%%%%%%%%%%%%%%%%%%%%%%%%%%%%%%%%%%%%%%%%
The ability to	control preparation and evolution of states of quantum systems
opens new horizons in  experimental physics (e.g. tests of fundamental
concepts of quantum mechanics) as well as in potential technical applications
(such as quantum information processing).
Recent advances in quantum optics (micromasers, cavity QED) \cite{Brune 1996}
and atomic physics (dynamics of trapped ions) \cite{Monroe 1995}
have demonstrated that a macroscopic observer can
effectively control dynamics as well as to  perform a complete measurement
of states  of microscopic quantum systems. In particular, preparation
of an {\it arbitrary} quantum states of a {\it single} mode
electromagnetic field
in micromasers and 1-dimensional vibrational motion of trapped ions have
been discussed in literature
\cite{special,Vogel 1993,Parkins 1993,Cirac 1993,Vogel 1995,Law 1996,Monroe 1996}. 
Experimental realizations
of highly non-classical states, such as Fock states, squeezed states
or Schr\"{o}dinger-cat states of these {\it single}-mode bosonic fields
have been reported \cite{Brune 1996,Monroe 1996}.
The ability to synthesize an arbitrary motional state is
a key prerequisite for a quantum measurement of an {\it arbitrary}
motional observable of a trapped ion as proposed by Gardiner,
Cirac and Zoller \cite{Gardiner 1997}.
These authors have generalized
the 1-D synthesis of motional states to 2-D and higher dimensions.
In spite of the conceptual elegance, the method proposed
by Gardiner, Cirac and Zoller is difficult to implement in general. The
problem is that the number of laser operations required for a preparation
of a two-mode target state $|\Psi_{target}\rangle$ of the form
\be
|\Psi_{target}\rangle =
\sum_{m=0}^{M_{max}}\sum_{n=0}^{N_{max}} Q_{mn}|m,n\rangle.
\label{st1}
\ee
depends {\it exponentially} on the dimensionality of the subspace of the
Fock space in which the target state is embedded.
In particular, if
$N_{max}=M_{max}$ then the number of necessary operations is proportional
to $2 M_{max} \times 2^{M_{max}}$. This exponential dependence restricts
applicability of the proposed method (see below).
A novel approach which overcomes this obstacle was introduces
very recently by Kneer and Law \cite{Kneer 1998}. The authors have considered
a photon--number--dependent interaction which is induced by a detuned
standing--wave laser field in $y$ direction. It keeps on resonance only
transitions between Fock states $|m,n\rangle$ with a fixed number $n=n_y$
of trap quanta in $y$ direction while applying sequentially 
the 1-D preparation scheme \cite{Law 1996} 
in the $x$ direction for each particular $n=n_y$.

In this paper we propose an alternative universal algorithm for
a construction (synthesis) of an {\it arbitrary}  quantum state of
the two-mode bosonic field which can be straightforwardly generalized
for the preparation of an arbitrary multi-mode system.
In our algorithm the number of operations required for a preparation
of the two-mode state (\ref{st1}) grows only polynomially as a function
of $N_{max}$ and $M_{max}$. In particular, if $N_{max}=M_{max}$, then
the number of operations is proportional to $8 M_{max}^2$.

To make our discussion as close as possible
to experimental realization we consider the preparation of
the two-dimensional vibrational motion of a trapped ion \cite{Gou 1996}.
The choice of the system is motivated by the fact that dissipative effects
can be significantly suppressed in  ion traps which is important for
a {\it deterministic} engineering of quantum states.
Moreover, the quantized vibration motion of a trapped ion
can be effectively controlled by a proper sequence of laser pulses tuned
either to the atomic electronic transition or to resolved vibrational
sidebands \cite{Monroe 1995,Monroe 1996}.

The simplest quantum state preparation is represented by a simple
{\it unitary}  evolution of an	input state of the system
governed by  a specific (generally, nonlinear) Hamiltonian.
Obviously, the fixed Hamiltonian restricts
the family of target states which can be ``generated'' from available inputs.
Another way how to prepare states of the system of interest (e.g,
single-mode bosonic field) is to consider quantum interaction between
this system and an adjoint quantum system (e.g. fermions). Dynamics
of these two systems is governed by a specific interaction Hamiltonian.
The desired quantum state engineering
is then achieved by  {\it conditional} measurements performed on  the adjoint
quantum system \cite{Vogel 1993}.
This {\it non-unitary} selection of specific quantum
trajectories  allows to synthesize essentially all quantum states of the
system under consideration. But there is prize to pay - the probability
of the outcome of the given conditional process can be extremely small.
To overcome this problem one may consider a sequence of interactions
between the original and the adjoint systems. These interactions (channels)
are  governed by different Hamiltonians with
just one channel  turned on at a given time.
Coupling constants in these Hamiltonians and times of duration
of given interactions are in this case {\it free}  parameters which
can be appropriately {\it tuned} so that at the output the system of interest
is disentangled from the adjoint system (see below) and is prepared in
the desired state. This approach has been recently utilized by
Law and Eberly \cite{Law 1996} who have shown that
by a suitable switching between
two channels of the atom-field interaction one can generate an arbitrary
state of the single-mode cavity field.
It is important to note that quantum states of 1-D
vibrational motion of trapped ions
are experimentally  created also via a sequence of laser pulses
tuned either to an electronic transition or an appropriate
vibrational sideband \cite{Monroe 1996}, that is, sequential switching of
different interaction channels is used.
Generalizations of the 1-D vibrational quantum state synthesis
to 2-D and more dimensions have been discussed recently
\cite{Gardiner 1997,Kneer 1998}.
In what follows we introduce an {\em universal} scheme which enables
deterministic preparation of an {\em arbitrary} two--mode target state
via an appropriate switching between
laser stimulated Raman transitions described by single and two--mode
interaction Hamiltonians.
The method can be generalized, e.g., for a trapped ion in a 3-D trap potential
and other multi-mode bosonic systems. Even for higher dimensions
the number of preparation operations scales polynomially with
the dimensionality of the subspace of the
Fock space in which the target state is embedded.

The paper is organized
as follows: in Section II we introduce our algorithm for 2-D quantum--state
synthesis. In Section III we discuss a physical realization of the
preparation scheme. In Section IV we 
examine the stability of our algorithm with respect to a technical noise.
Fidelity between outputs  and  specific target states is evaluated.
We finish our paper with conclusions.

%%%%%%%%%%%%%%%%%%%%%%	SYNTHESIS  %%%%%%%%%%%%%%%%%%%%%%%%%%%%%%%%%%%%%%
\section{Quantum-state synthesis}
%%%%%%%%%%%%%%%%%%%%%%%%%%%%%%%%%%%%%%%%%%%%%%%%%%%%%%%%%%%%%%%%%%%%%
Without any loss of generality let us assume that we want to
generate an arbitrary two--mode state
given as a finite superposition of number (Fock) states (\ref{st1}).
To describe the preparation algorithm we firstly split
the whole Hilbert space into an appropriate
set of finite-dimensional subspaces labeled by a specific quantum number.
Then we introduce two sorts of interaction channels. The first set
of Hamiltonians will be responsible for dynamics within a given subspace;
while the second one will realize a ``transfer'' of probability amplitudes
between different subspaces.
In particular, let us consider vibrational states of a trapped ion,
confined in 2-D harmonic potential.
Excitations of two vibrational modes are described by  creation and
annihilation operators $\hat{a}_{\mu}$ and $\hat{a}_{\mu}^\dagger$
($\mu=x,y$).
The Hilbert space ${\cal H}_{vib}$ of all two--mode vibrational states
can be divided to subspaces with constant total number of vibrational
quanta, i.e.,
${\cal H}_{vib}=\oplus {\cal H}_{J}$, $J=0,1,\ldots,\infty$,
where ${\cal H}_{J}$ is spanned by two--mode Fock states
$|J,0\rangle, |J-1,1\rangle, \ldots, |0,J\rangle$.
The subspaces ${\cal H}_{J}$ are dynamically independent
(invariant) for those Hamiltonians which do not change
the total number of vibrational quanta
$\hat{J}=\hat{a}_{x}^\dagger\hat{a}_{x}+\hat{a}_{y}^\dagger\hat{a}_{y}$.
However, to realize
controlled manipulations with  states within a given subspace  ${\cal H}_{J}$
we need an interaction of the 2-D bosonic field with another
quantum system. In the case of trapped ions it
is natural to assume  the coupling to internal energy levels of ions.
In particular, we use three  internal electronic states $|i\rangle$
(in $\Lambda$ or $\Xi$ configurations)
with energies $\hbar\omega_i$ ($i=a, b, c$)
which form the basis of the Hilbert subspace ${\cal H}_{in}$.
Due to quantum interaction between the vibrational and internal degrees
of freedom the total state vector of the composed system with the given
maximum number of vibrational quanta $J_{max}=M_{max}+N_{max}$
at time $t$ reads
\be
|\Psi(t)\rangle &=&  \sum_{m=0}^{M_{max}}\sum_{n=0}^{N_{max}} \sum_{i=a,b,c}
		Q_{m,n;i}(t) |m,n\rangle \otimes |i\rangle \nonumber \\
		&=&  \sum_{J=0}^{J_{max}}\sum_{k=0}^{J} \sum_{i=a,b,c}
		Q_{k,J-k;i}(t) |k,J-k\rangle \otimes |i\rangle
,
\label{st2}
\ee
which reflects quantum-mechanical entanglement between the two subsystems.
We assume that	initially the ion is in an internal state $|a\rangle$
and  the vibrational motion is cooled to the ground state
$|0,0\rangle$. That is, the initial state vector of the composed system
$|\Psi(t=0)\rangle = |0,0\rangle \otimes |a\rangle$ is factorized.
Our task is to find an unitary evolution $\hat{U}^\dagger$,
such that at some time $t=T$ the
state vector (\ref{st2}) can again be factorized, while the
vibrational state of the ion is described by the target vector (\ref{st1}),
i.e. $|\Psi(t=T)\rangle=\hat{U}^\dagger |0,0\rangle \otimes |a\rangle
= |\Psi_{target} \rangle \otimes |a\rangle$.

In what follows we prove that the unitary transformation $\hat{U}$
can be represented by a sequence of five ``elementary''
unitary transformations $\hat{U}^{(p)}$ ($p=1,...,5$)
which act in the product Hilbert space
${\cal H}_{vib} \otimes {\cal H}_{in}$ and
which correspond to five interaction channels associated with
interaction Hamiltonians $\hat{H}^{(p)}$:
\end{multicols}
\vspace{-0.5cm}
\noindent\rule{0.5\textwidth}{0.4pt}\rule{0.4pt}{\baselineskip}
%\widetext
\be
\hat{U} = &&
\hat{A}_{0} \{ \prod_{J=1}^{J_{max}}\hat{C}_{J}\hat{B}_{J-1}\hat{A}_{J} \};
\qquad
\hat{A}_{J}=  \hat{U}^{(1)}_{|J,0,b\rangle } \hat{U}^{(3)}_{|J-1,1,a\rangle}
  \ldots   \hat{U}^{(1)}_{|2,J-2,b\rangle} \hat{U}^{(3)}_{|1,J-1,a\rangle}
	   \hat{U}^{(1)}_{|1,J-1,b\rangle} \hat{U}^{(3)}_{|0,J,a\rangle};
\nonumber
\\ 
&& \hat{B}_{J} = \hat{U}^{(2)}_{|J,0,c\rangle }
      \hat{U}^{(4)}_{|J-1,1,b\rangle} \hat{U}^{(2)}_{|J-1,1,c\rangle}
       \ldots \hat{U}^{(4)}_{|1,J-1,b\rangle} \hat{U}^{(2)}_{|1,J-1,c\rangle}
            \hat{U}^{(4)}_{|0,J,b\rangle} \hat{U}^{(2)}_{|0,J,c\rangle};
\qquad
\hat{C}_{J}=\hat{U}^{(5)}_{|J,0,b\rangle}
\label{st3}
\ee

%%%%%%%%%%%%%%%%%%%%%% FIGURE 1 %%%%%%%%%%%%%%%%%%%%%%%%%%%%%%%%%
\begin{figure}
%% Fig.1
\centerline{\epsfig{height=11.0cm,file=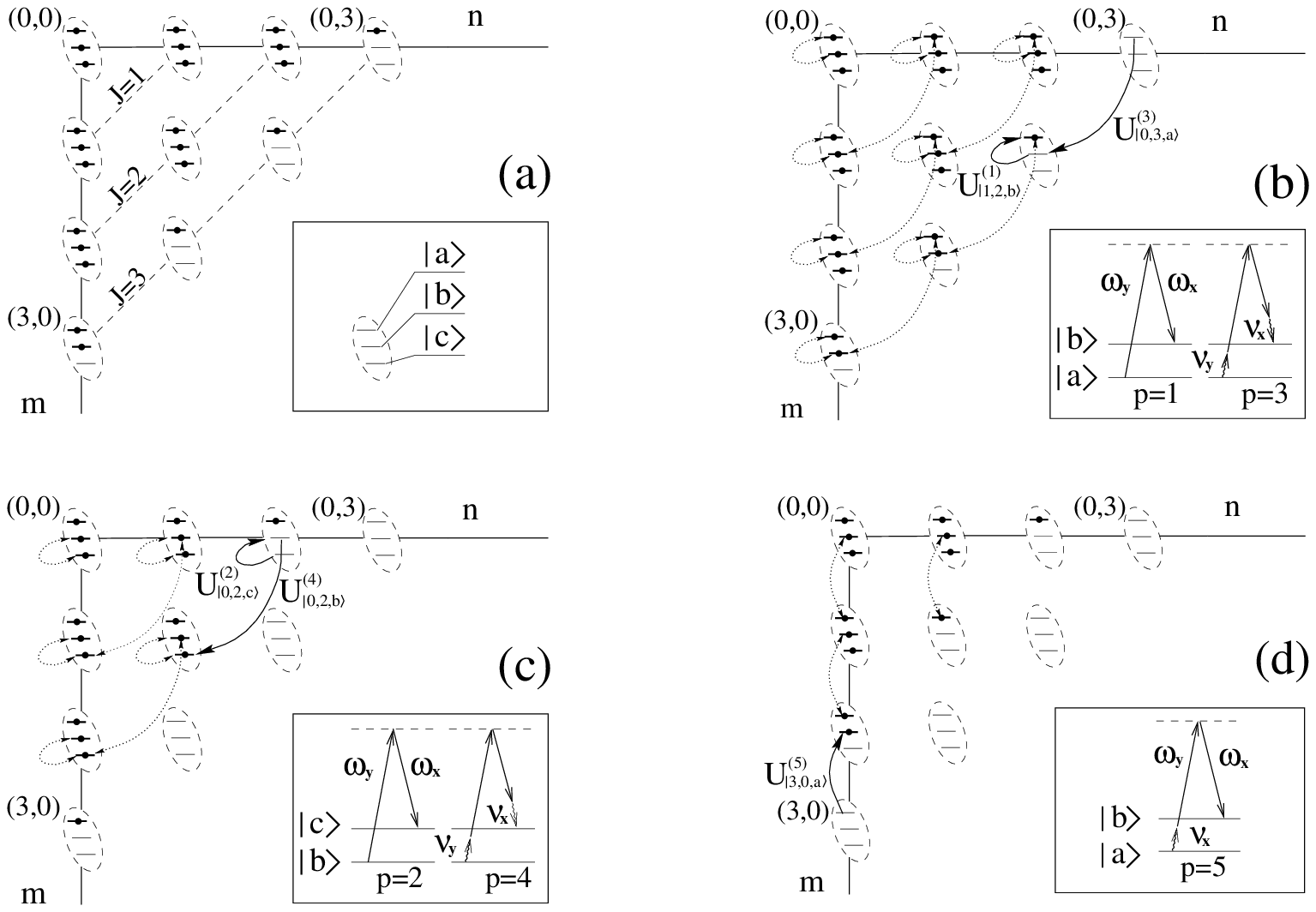}}
%\caption{}
%%Warning: increase the buffer size}
\end{figure}
%%%%%%%%%%%%%%begin figure caption
%\def\cap1
{\small FIG. 1.
Recursive ``de-evolution'' algorithm:
(a) Vectors  $|m,n\rangle \otimes |i\rangle$ are represented
as points of a
lattice $m,n$ with internal levels $i=a,b,c$ shown in ``ovals''.
Components which enter (at this stage of ``de-evolution'')
the state vector (\ref{st2}) are indicated by $\bullet$.
Dashed lines connect  basis vectors from subspaces
with constant number of trap quanta $J$.
(b) The action of the operator $\hat{A}_{\tilde{J}}$ for $\tilde{J}=3$.
The ``elementary'' transformation
$\hat{U}^{(3)}_{|0,\tilde{J},a\rangle}$ transfers {\em completely}
the population of the state $|0,\tilde{J}\rangle \otimes |a\rangle$
to the state $|1,\tilde{J}-1\rangle \otimes |b\rangle$ as indicated
by solid arrow. The atomic transition
scheme of the corresponding stimulated Raman process
$p=3$ is shown in the inset.
Next $\hat{U}^{(1)}_{|1,\tilde{J}-1,b\rangle}$ transfers the
population from $|1,\tilde{J}-1\rangle \otimes |b\rangle$
to $|1,\tilde{J}-1\rangle \otimes |a\rangle$ by means of the interaction
channel $p=1$. The dashed arrows indicate simultaneous
transitions which lead to a change of the state vector
(\ref{st2}). The application of the  sequence
$\hat{A}_{\tilde{J}}$ ``shrinks'' the whole  population from the given
subspace ${\cal H}_{\tilde{J}} \otimes {\cal H}_{in}$
to the component state $|1,\tilde{J}-1\rangle \otimes |a\rangle$.
(c) Specific preparation of the subspace
${\cal H}_{\tilde{J}-1} \otimes {\cal H}_{in}$ by
the operator $\hat{B}_{\tilde{J}-1}$.
The operator $\hat{U}^{(2)}_{|0,\tilde{J}-1,c\rangle}$ transfers
the population of the state $|0,\tilde{J}-1 \rangle \otimes |c\rangle$
to $|0,\tilde{J}-1\rangle \otimes |b\rangle$,
while $\hat{U}^{(4)}_{|0,\tilde{J}-1,b\rangle}$ transfers
the population from $|0,\tilde{J}-1 \rangle \otimes |b\rangle$
to $|1,\tilde{J}-2\rangle \otimes |b\rangle$.
The inset shows the schemes of the utilized stimulated Raman transitions
$p=2,4$. Finally, the sequence $\hat{B}_{\tilde{J}-1}$ leaves
on ${\cal H}_{\tilde{J}-1} \otimes {\cal H}_{in}$ only
the component state $|\tilde{J}-1,0\rangle \otimes |b\rangle$
and those states
with the internal level $|a\rangle$ contributing to (\ref{st2}).
(d) The operator $\hat{C}_{\tilde{J}}=\hat{U}^{(5)}_{|\tilde{J},0,a\rangle}$
transfers  the population of the component state
$|\tilde{J},0\rangle \otimes |a\rangle$ to the state
$|\tilde{J}-1,0\rangle \otimes |b\rangle$.
The procedures (b)-(c) are recursively repeated to ``de-evolve''
the initial state
$|\psi_{target}\rangle\otimes |a\rangle$ into $|0,0\rangle\otimes
|a\rangle$.
}
%%%%%%%%%%%%%end of fig. caption
%%%%%%%%%%%%%%%%%%%%%%%% FIGURE 1 %%%%%%%%%%%%%%%%%%%%%%%%%%%%%%%%%
\begin{multicols}{2}

The operators $\hat{A}_{J}$, $\hat{B}_{J}$, $\hat{C}_{J}$
are build up from ``elementary'' unitary transformations
$\hat{U}^{(p)}_{|k,J-k,i\rangle}$ where superscript $p$ determines
the interaction channel and subscript indicates that
parameters of the interaction are adjusted to transfer {\em completely} the
current population of the component state $|k,J-k\rangle \otimes |i\rangle$
to the ``neighboring''  state
according to the given type of interaction (see below).

The set of the ``elementary'' unitary transformations $\hat{U}^{(p)}$
is sufficient (even though not unique) to evolve the target state
(\ref{st1}) into the vacuum  and vice versa. To specify them we will
analyze the ``de-evolution'' of the target state into the vacuum state.
We will perform this de-evolution via a systematic ``transfer'' of population
from state vectors with higher to smaller number
of vibrational quanta. That is, we choose the operators $\hat{U}^{(p)}$
so that at a given step of the unitary de-evolution a particular
probability amplitude  $Q_{k,J-k;i}$ is made equal to zero.
In Fig.1. we visualize the
``direction'' of the action of operators $U^{(p)}_{|k,J-k,i\rangle}$.
Namely, in Fig.1a  we represent  state vectors
$|m,n\rangle \otimes |i\rangle$
as ``ovals'' located at the points of the lattice marked by parameters
$m,n$ associated with vibrational state of an ion, while its internal
state represented by a simple energy level diagram
[population of a given internal level is indicated  by a bullet ($\bullet$)].
Dashed lines indicate the subspaces
${\cal H}_{J} \otimes {\cal H}_{in}$
labeled by a constant number of trap quanta $J$.

Let us start the ``de-evolution'' procedure of the target state
(\ref{st1}) into vacuum.
After time $\tau=T-t$ the vector describing  the vibrational
and internal state of the ion is given by Eq.(\ref{st2}).
We assume
that the de-evolution is performed in such way that at this moment
the state vector (\ref{st2})  is composed only	of state vectors
from subspaces	${\cal H}_{J} \otimes {\cal H}_{in}$ with $J\le \tilde{J}$
($\le J_{max}$). Now we apply three
sequences of operations described the operators
$\hat{A}_{\tilde J}$, $\hat{B}_{{\tilde J}-1}$,  and $\hat{C}_{\tilde J}$,
respectively [see Eq.(\ref{st3})].

The action of the operator $\hat{A}_J$	is illustrated in Fig.~1b.
It describes a process in which the first
``elementary'' transformation
$\hat{U}^{(3)}_{|0,\tilde{J},a\rangle}$ transfers {\em completely}
the population of the state $|0,\tilde{J}\rangle \otimes |a\rangle$
to the state $|1,\tilde{J}-1\rangle \otimes |b\rangle$
(see the  solid arrow in Fig.1b; dashed arrows indicate simultaneous
transitions which are not controlled at the given stage).
This process
can be realized by irradiating the ion with two external laser fields
with tunable frequencies $\omega_x$ and $\omega_y$ in $x$ and $y$
directions. Adjusting the resonance conditions
$\omega_{y}-\omega_{x}=\omega_b-\omega_a+\nu_{x}-\nu_{y}$
($\equiv\omega_{3}$)
the stimulated Raman transition [see the inset in Fig.1b, $p=3$]
can be described in the Lamb--Dicke regime
by the effective interaction Hamiltonian 
(we discuss physical conditions under which this and all subsequent
Hamiltonians can be justified in Section III):
\be
\hat{H}_{3}= g_{3} \hat{a}_{x}^\dagger\hat{a}_{y} |b\rangle\langle a|
e^{-i\omega_{3}t} +
g_{3}^{*} \hat{a}_{x}\hat{a}_{y}^\dagger |a\rangle\langle b|
e^{i\omega_{3}t}
\label{ha3}
;\ee
with the corresponding time evolution operator
$\hat{U}^{(3)}=\exp(-{\rm i}\hat{H}^{(3)}t)$ setting $\hbar=1$.
If the interaction constant $g_3=|g_3| {\rm e}^{{\rm i}\theta_3}$
and the duration of the interaction $t$ are chosen to fulfill the condition
\be
{\rm i} e^{{\rm i}\theta_{3}}
Q_{k,J-k;a}\cos(|g_{3}|t\sqrt{(k+1)(J-k)})      &+& \nonumber \\
Q_{k+1,J-k-1;b} \sin(|g_{3}|t\sqrt{(k+1)(J-k)}) &=& 0,
\label{c1}
\ee
then the population of the component state $|k,J-k\rangle \otimes |a\rangle$
is completely transferred to the state $|k+1,J-k-1\rangle \otimes
|b\rangle$.
(for instance, in the situation described by Fig.1b
the probability amplitude  $Q_{0,\tilde{J};a}$ becomes equal to zero).
Once this is done, then the  transformation
$\hat{U}^{(1)}_{|1,\tilde{J}-1,b\rangle}$ is turned on. In this process
the population of the component state
$|1,\tilde{J}-1\rangle \otimes |b\rangle$ is {\em completely} transferred
to the state $|1,\tilde{J}-1\rangle \otimes |a\rangle$.
The corresponding interaction channel is described by the
Hamiltonian
\be
\hat{H}_{1}=g_{1} |b\rangle\langle a| e^{-i\omega_{1}t}
+ g_{1}^{*} |b\rangle\langle a| e^{i\omega_{1}t}
\label{ha1}
.\ee
Here we assume lasers to be tuned to the electronic transition, i.e.,
$\omega_{y}-\omega_{x}=\omega_b-\omega_a$ ($\equiv \omega_1$).
To cancel the term $Q_{k,J-k;b}$ in the state vector (\ref{st2}),
the interaction constants ($|g_1|t$, $\theta_1$) have to be chosen
to satisfy the condition
\be
Q_{k,J-k;a}\sin(|g_{1}|t)+
{\rm i} e^{-{\rm i}\theta_{1}}  Q_{k,J-k;b}\cos(|g_{1}|t)=0.
\label{c2}
\ee
In particular, in Fig.~1b the operator
$\hat{U}^{(1)}_{|1,\tilde{J}-1,a\rangle}$
cancels $Q_{1,\tilde{J}-1;b}$ for $\tilde{J}=3$.
The successive action of the
``elementary'' transformations $\hat{U}^{(3)}$ and $\hat{U}^{(1)}$
which form the operator $\hat{A}_{\tilde{J}}$ [see Eq.(\ref{st3})]
finally ``shrinks'' the population of the subspace
${\cal H}_{\tilde{J}} \otimes {\cal H}_{in}$ to  a single state
$|\tilde{J},0\rangle \otimes |a\rangle$.

At this stage we start
the process of cancellation of the contribution
of  component states $|k,\tilde{J}-1-k\rangle \otimes |b\rangle$
and $|k,\tilde{J}-1-k\rangle \otimes |c\rangle$
in the ``neighboring'' subspace
${\cal H}_{\tilde{J}-1} \otimes {\cal H}_{in}$ (see Fig.1c).
This intermediate procedure is required to prevent a reverse transfer
of population from ${\cal H}_{\tilde{J}-1} \otimes {\cal H}_{in}$
to ${\cal H}_{\tilde{J}} \otimes {\cal H}_{in}$ (see below and Fig. 1d).
For this purpose the operator $\hat{B}_{\tilde{J}-1}$ is constructed
from ``elementary'' operations $\hat{U}^{(2)}$ and $\hat{U}^{(4)}$
[see Eq.(\ref{st3})]. Namely, the operator
$\hat{U}^{(2)}_{|0,\tilde{J}-1,c\rangle}$ describes the transfer
of population
of the state $|0,\tilde{J}-1 \rangle \otimes |c\rangle$
to $|0,\tilde{J}-1\rangle \otimes |b\rangle$. This transfer can be achieved
with the help of
lasers pulses tuned to the electronic transition
between the levels $|b\rangle$ and $|c\rangle$. The corresponding
interaction Hamiltonian $\hat{H}^{(2)}$ and the resonance condition
are analogous to those for $\hat{H}^{(1)}$ (\ref{ha1}) [we have to replace
only  $b\to c$ and $a\to b$]. Further the operation
$\hat{U}^{(4)}_{|0,\tilde{J}-1,b\rangle}$ cancels a contribution
of the state $|0,\tilde{J}-1\rangle \otimes |b\rangle$
to the state vector (\ref{st2}). This interaction channel is
described by  the Hamiltonian $\hat{H}^{(4)}$
which is obtained from (\ref{ha3}) by the  substitution $b\to c$ and $a\to b$.
We see that
the operator $\hat{B}_{\tilde{J}-1}$ acts like
$\hat{A}_{\tilde{J}}$  but instead of the stimulated
Raman processes between $a\leftrightarrow b$
the transitions $b\leftrightarrow c$ are utilized.
As the result of the action of the operator  $\hat{B}_{\tilde{J}-1}$
in the subspace  ${\cal H}_{\tilde{J}-1} \otimes {\cal H}_{in}$ only
the component states with the internal level $|a\rangle$
and the state $|\tilde{J}-1,0\rangle \otimes |b\rangle$
have non-zero amplitudes.

After the action of the operators $\hat{A}_{\tilde{J}}$ and
$\hat{B}_{\tilde{J}-1}$
the unitary operator $\hat{C}_{\tilde{J}}$ is utilized
to transfer the population of the component state
$|\tilde{J},0\rangle \otimes |a\rangle$ to the state
$|\tilde{J}-1,0\rangle \otimes |b\rangle$ (see Fig.~1d for
$\tilde{J}=3$).
The transformation
$\hat{U}^{(5)}_{|\tilde{J},0,a\rangle}$ which performs transitions
between subspaces with the number of trap quanta differed by one
is realized by the process described by
a  single--mode interaction Hamiltonian
\be
\hat{H}_{5}=g_{5} \hat{a}_{x}|b\rangle\langle a| \mbox{e}^{-i\omega_{5}t} +
g_{5}^* \hat{a}_{x}^\dagger |a\rangle\langle b| \mbox{e}^{+i\omega_{5}t}
\label{ha5}
\ee
with the resonance condition
$\omega_{y}-\omega_{x}= \omega_b-\omega_a -\nu_{x}$ ($\equiv \omega_5$).
The parameters of this interaction channel are determined by
the constraint
\be
{\rm i} e^{{\rm i}\theta_{5}} Q_{J,0;a}\cos(|g_{5}|t\sqrt{J}) +
Q_{J-1,0;b} \sin(|g_{5}|t\sqrt{J}) &=& 0,
\label{c3}
\ee
for $J=\tilde{J}$.

As the result of the action of the operators
$\hat{A}_{\tilde J}$, $\hat{B}_{{\tilde J}-1}$,  and $\hat{C}_{\tilde J}$,
all  coefficients $Q_{k,\tilde{J}-k;i}$ in (\ref{st2})
are equal to zero.
Moreover, the situation before (Fig.~1a) and after (Fig.~1d) the
action of these  operators
is the same, except  we just ``moved'' from the subspace
${\cal H}_{\tilde{J}} \otimes {\cal H}_{in}$
to ${\cal H}_{\tilde{J}-1} \otimes {\cal H}_{in}$ with the number
of trap quanta decreased by one.
This means that the procedure can be {\em recursively} repeated.
The given solution of the ``de-evolution'' gives immediately the recipe
for the creation of the target vibrational state from the two--mode vacuum
(with electronic level $|a\rangle$): We have to change by $\pi$
the phase shift between external laser fields and repeat the sequence
in the opposite order, i.e. we apply $\hat{U}^\dagger$ on
$|0,0\rangle\otimes |a\rangle$.
In this way we can synthesize an arbitrary state.
 The number of
``elementary'' operations involved in this process
is proportional to $2 J_{max}^2$, which is important for an experimental
realization of the proposed scheme, i.e., the number of necessary operations
increases only polynomially with the increase of the size of the Hilbert
space in which the target state is embedded.

In the following section a physical implementation of the preparation scheme
is discussed in more details.

%%%%%%%%%%%%%%%%%%%%%%%%%%%%%%%%%%%%%%%%%%%%%%%%%%%%%%%%%%%%%%%%%%%%%%%%%%%%%
\section{Realization of interaction channels}
%%%%%%%%%%%%%%%%%%%%%%%%%%%%%%%%%%%%%%%%%%%%%%%%%%%%%%%%%%%%%%%%%%%%%%%%%%%%%

The Hamiltonians which are eligible for the synthesis of two--mode bosonic
states has been discussed recently by several authors
\cite{Gardiner 1997,Kneer 1998,Gou 1996,Steinbach 1997}.
Some of the proposals are based on laser stimulated dipole transitions
\cite{Gardiner 1997} and phonon-number-dependent interaction
via detuned standing--wave \cite{Kneer 1998}.
For our purposes we have utilized the stimulated Raman processes
discusses in detail by {Steinbach, Twamley} and {Knight}
\cite{Steinbach 1997}. In this section following Ref.\cite{Steinbach 1997}
 we briefly derive the Hamiltonians
which are used in our algorithm
[see Eqs.(\ref{ha3}),(\ref{ha1}),(\ref{ha5})] 
and discuss the range of their applicability.
Let us consider a trapped ion confined in a 2D harmonic potential
characterized by the trap frequencies $\nu_x$ and $\nu_y$
in two orthogonal directions $x$ and $y$.
The ion is irradiated along the $x$ and $y$ axes by two external laser
fields with frequencies $\omega_x$, $\omega_y$ and wave vectors
$k_x$, $k_y$.
The laser fields stimulate Raman transitions between two internal energy
levels $|a\rangle$ and $|b\rangle$ via an auxiliary electronic
level $|r\rangle$ which is far off resonance.
For concreteness,
we consider $\Lambda$-configuration with the upper
level $|r\rangle$ as outlined in the inset of Fig.~1b.
The interaction Hamiltonian for the system under consideration can be
written in the dipole and rotating--wave approximation
(RWA at laser frequencies) in the form:
\be
\hat{H}_{int} &=&
 g_{x}^* \mbox{e}^{i(k_x\hat{x}-\omega_{x}t)} |r\rangle \langle b| +
 g_{x} \mbox{e}^{-i(k_x\hat{x}-\omega_{x}t)} |b\rangle \langle r|
\nonumber \\
&+&
  g_{y}^* \mbox{e}^{i(k_y\hat{y}-\omega_{y})t}  |r\rangle \langle a| +
  g_{y} \mbox{e}^{-i(k_y\hat{y}-\omega_{y})t}  |a\rangle \langle r|
.\label{ap1} \ee
The coupling constant $g_x$ ($g_y$) is  proportional to the intensity
of the laser field in $x$ ($y$) direction and the dipole moment of
the electronic transition $|b\rangle \leftrightarrow  |r\rangle$
($|a\rangle \leftrightarrow  |r\rangle$).
The upper off-resonant level $|r\rangle$ can be adiabatically eliminated
provided that $\Delta_x,\Delta_y \gg g_a,g_b,|\Delta_x-\Delta_y|$, where
laser detunings for dipole transitions $|b\rangle \leftrightarrow  |r\rangle$
and $|a\rangle \leftrightarrow |r\rangle$
are denoted as $\Delta_x=(\omega_r-\omega_b)-\omega_x$ and
$\Delta_y=(\omega_r-\omega_a)-\omega_y$, respectively.
After adiabatic elimination the effective interaction Hamiltonians
for the stimulated Raman transition $|a\rangle \leftrightarrow  |b\rangle$
reads \cite{Steinbach 1997}
\be
\hat{H}_{int}^{(eff)} &=&
  g^* \mbox{e}^{-i(\omega_{y}-\omega_x)t}
 \hat{D}_{x}(-i\epsilon_x) \hat{D}_{y}(i\epsilon_y) |b\rangle \langle a|
\nonumber \\
&+&   g~ \mbox{e}^{i(\omega_y-\omega_x)t}
 \hat{D}_{x}(i\epsilon_x) \hat{D}_{y}(-i\epsilon_y) |a\rangle \langle b|
.\label{ap2} \ee
Here $\hat{D}_{q}(i\epsilon_q)=
\mbox{e}^{i\epsilon_q(\hat{a}^{\dagger}_{q}+\hat{a}_{q})}=
\mbox{e}^{i k_q\hat{q}}$
is the displacement operator ($q=x,y$);
the Lamb--Dicke parameter $\epsilon_q$ is defined as
$\epsilon_q^2= \hbar^2 k_q^2/(2m\hbar\nu_q)$ and
the effective interaction constant
$g=g_x^* g_y (\frac{1}{ \Delta_x}+\frac{1}{ \Delta_y})$.
Further, we assume that the energies of the electronic levels
$|a\rangle$ and $|b\rangle$ are redefined to include
Stark shifts due to the adiabatic elimination of the off-resonant 
energy level $|r\rangle$ \cite{Steinbach 1997}.

In the interaction picture the effective interaction Hamiltonian
$\tilde{\hat{H}}_{int}^{(eff)}=
\mbox{e}^{i\hat{H}_0 t} \hat{H}_{int}^{(eff)} \mbox{e}^{-i\hat{H}_0 t}$
can be expressed as
[the free Hamiltonian $\hat{H}_0$ induces transformations
$\hat{a}_q\to\hat{a}_q \mbox{e}^{-i\nu_q t}$,
$|a\rangle\langle b|\to |a\rangle\langle
b|\mbox{e}^{-i(\omega_b-\omega_a)t}$]
\end{multicols}
\vspace{-0.5cm}
\noindent\rule{0.5\textwidth}{0.4pt}\rule{0.4pt}{\baselineskip}
%\widetext
\be
\tilde{\hat{H}}_{int}^{(eff)} &=&
  g^* \mbox{e}^{-\frac{\epsilon_x^2+\epsilon_y^2}{2}}
\sum_{m,k,l,n}
\frac{(-i\epsilon_x)^{k+m} (i\epsilon_y)^{l+n}}{k!l!m!n!}
\mbox{e}^{-it[\Delta+(k-m)\nu_x +(n-l)\nu_y]}
\hat{a}_x^{\dagger m} \hat{a}_x^k \hat{a}_y^{\dagger l} \hat{a}_y^n
|b\rangle \langle a|  + h.c.
\label{ap3} \ee
\begin{multicols}{2}
The resonant terms in the expansion (\ref{ap3}) which contribute
dominantly to the resulting effective Hamiltonian can be selected by
an appropriate choice of laser frequencies.
If off--resonant processes are oscillating with sufficiently high frequencies
they can be eliminated applying the second RWA at trap frequencies.

In particular, tuning lasers to the first red sidebands
$\Delta\equiv\Delta_x-\Delta_y=\nu_x-\nu_y$ with {\em incommensurate}
trap frequencies only the resonant terms
with $k-m=1$ and $l-n=1$ are retained in the expansion (\ref{ap3}):
\be
\tilde{\hat{H}}_{int}^{(3)} &=&
  g^* \epsilon_x\epsilon_y \hat{a}_x^\dagger
\hat{\cal F}(\hat{a}_x^\dagger\hat{a}_x,\hat{a}_y^\dagger\hat{a}_y)
\hat{a}_y |b\rangle \langle a| + h.c. \label{ap4} 
\ee
where
\be
\hat{\cal F}&=&\mbox{e}^{-\frac{\epsilon_x^2+\epsilon_y^2}{2}}
\sum_{k,l} \frac{(-1)^{k+l} \epsilon_x^{2k} \epsilon_y^{2l}}{(k+1)!k!(l+1)!l!}
\hat{a}_x^{\dagger k} \hat{a}_x^k \hat{a}_y^{\dagger l} \hat{a}_y^l
.\ee
In the Lamb--Dicke regime $\epsilon_x,\epsilon_y \ll 1$ the operator
$\hat{\cal F}$ is close to the unity operator and the Hamiltonian
(\ref{ap4}) which is written in the interaction picture acquires in
the Schr\"odinger picture exactly the form of the two--mode Hamiltonian
$\hat{H}_3$ (\ref{ha3}).
With proper laser tunings we can design within the Lamb--Dicke limit
also remaining interaction Hamiltonians required for the quantum state 
synthesis described in Section II.
In particular, retaining only resonant terms in (\ref{ap3}) 
for the laser-stimulated Raman process with $\Delta=0$ 
the interaction Hamiltonian $\hat{H}_1$ (\ref{ha1}) is obtained in
the Schr\"odinger picture. The process with $\Delta=-\nu_x$
is described by the one--mode interaction Hamiltonian $\hat{H}_5$ (\ref{ha5}).
Let us also notice that beyond the  Lamb--Dicke limit 
this process is analogous to a 
nonlinear Jaynes--Cummings dynamics discussed by Vogel and de Matos Filho
\cite{Vogel 1995}.

Considered approximations impose limitations on the applicability
of the interaction Hamiltonians. Driven electronic transition (\ref{ha1})
and one--mode interaction (\ref{ha5}) have been considered already
for the 1-D quantum state synthesis \cite{Law 1996}.
A new tool in our approach represents the two--mode interaction
Hamiltonian $\hat{H}_3$ (\ref{ha3}).
Comprehensive analysis of the limitations for (\ref{ha3})
was done in Ref.\cite{Steinbach 1997}.
It turns out that the most subtle point is the second RWA at trap
frequencies [see Eqs.(\ref{ap3}),(\ref{ap4})]
which imposes restrictions \cite{Steinbach 1997}
\be
|g| \epsilon_x\epsilon_y \max(N_{max},M_{max}) &\ll&  \min(\nu_x,\nu_y), 
\label{ap5} \\ 
\frac{\min(\nu_x,\nu_y)}{\max(\nu_x,\nu_y)} &\ge& 5
.\nonumber\ee
The trap anisotropy, i.e., {\em incommensurate} trap frequencies,
is required to avoid additional resonances in (\ref{ap3}).
Numerical simulations in Ref.\cite{Steinbach 1997} demonstrated
that it is experimentally feasible to operate the considered Hamiltonians
within the Lamb--Dicke limit.

It is worth noticing that already the nonlinear form (\ref{ap4})
corresponding to the two--mode interaction Hamiltonian (\ref{ha3})
outside of the Lamb--Dicke regime
allows us to adopt the proposed algorithm for quantum state synthesis.
The main difference consists in the form of  generalized Rabi frequencies.
In particular, the matrix element of the interaction Hamiltonian (\ref{ap4})
in the Lamb--Dicke regime $\epsilon_q\ll 1$
[i.e., (\ref{ha3}) in the interaction picture] reads
$\langle b,n-1,m+1|\tilde{\hat{H}}_{int}^{(3)}|m,n,a\rangle=
g_3 \sqrt{(m+1)n}$
while beyond the Lamb-Dicke regime the corresponding matrix element
is given as $g_3 \mbox{e}^{-\frac{\epsilon_x^2+\epsilon_y^2}{2}}
L_{m}^{1}(\epsilon_x^2) L_{n-1}^{1}(\epsilon_y^2)/\sqrt{(m+1)n}$
where $L^{1}_{m}$ is the associated Laguerre polynomial.
Inserting the ``nonlinear'' Rabi frequencies into 
Eqs.(\ref{c1}),(\ref{c2}),(\ref{c3}) which determine 
the choice of the interaction constants and switching times,
the same sequence of elementary transformations (\ref{st3}) could be 
applied for preparation of a given state 
even outside of the Lamb--Dicke regime. 
On the other hand, the conditions for the applicability of the second
RWA at trap frequencies to obtain (\ref{ap4}) from (\ref{ap3})
are not so transparent as restrictions for the RWA within 
the Lamb--Dicke regime (\ref{ap5}). 
This problem goes beyond the scope of the present paper in which we
demonstrates the potential of our algorithm in the Lamb-Dicke regime.

%%%%%%%%%%%%%%%%%%%%%%%%%%%%%%%%%%%%%%%%%%%%%%%%%%%%%%%%%%%%%%%%%%%
\section{Stability of synthesis}
%%%%%%%%%%%%%%%%%%%%%%%%%%%%%%%%%%%%%%%%%%%%%%%%%%%%%%%%%%%%%%%%%%%

Our algorithm works ideally when there are essentially
no dissipations in the
system. It is the reason why we have consider effectively dissipation-free
dynamics of trapped ions. Nevertheless, we have to stress that
 the ideal synthesis
of quantum motional states assumes also perfect control of interaction
constants and switching times of particular interaction channels.
Namely,  the values of $g_p t$ found via the ``de-evolution'' procedure
[i.e., solutions of Eqs.(\ref{c1},\ref{c2},\ref{c3})] have to be controlled
as precisely as possible.
In practice one cannot avoid some level of ``technical'' noise,
for example, due to imperfect timing of switching between
interaction channels.
Therefore in what follows  we will study the stability of the presented 
 algorithm
with respect to the ``technical'' noise.
This  noise is simulated as random fluctuations
of the ideal values $g_p t$. In particular, fluctuations are
equally distributed around the ideal (complex) values $g_p t$
[i.e., solutions of Eqs.(\ref{c1},\ref{c2},\ref{c3})]
within a fixed interval $(1+{\rm i})\delta$.
In general, the output state vectors  $|\Psi_{\delta}\rangle$ of the
composed system which are prepared in the presence of ``technical''
fluctuations take the form (\ref{st2}), i.e.,
the vibrational and  internal degrees of freedom are not disentangled.
The Fig.~2 shows the fidelity $f$ of the ``imperfect'' output states
$|\Psi_{\delta}\rangle$
with respect to
the desired state vector $|\Psi_{target}\rangle\otimes |a\rangle$.
The fidelity is defined as the averaged squared scalar product of particular
realizations $|\Psi_{\delta}\rangle$
with  $|\Psi_{target}\rangle\otimes |a\rangle$,
i.e.,
\be
f=\langle\!\langle |\langle\Psi_{\delta}|
\Psi_{target}\rangle|a\rangle|^2 \rangle\!\rangle_{\delta}.
\label{fid}
\ee
In our simulations we have performed averaging over 100 runs of
state-synthesis sequences. In these runs each value $q_p t$ associated
with a given elementary operation $\hat{U}^{(p)}$ acquires a random
fluctuation within the interval $(1+i)\delta$.
We have considered
two different target states,
two--mode cat--like state $|\Psi_{cat}\rangle={\cal N}_{cat}
(|\alpha\rangle |\alpha\rangle + |-\alpha\rangle |-\alpha\rangle)$
and two--mode correlated state
$|\Psi_{corr}\rangle=
{\rm e}^{-|\alpha|^2/2} \sum_{m} \frac{\alpha^m}{\sqrt{m!}} |m,m\rangle$,
with $\alpha =2$. For this value of the amplitude these two states
have approximately the same mean number of vibrational quanta
$\bar{n}= \bar{n}_x+\bar{n}_y \simeq 8.0$.
We have considered two cases when  $M_{max}=12$ and $M_{max}=20$
(here $N_{max}=M_{max}$).

Fig.~2 clearly indicates the fact, that larger  the value of $M_{max}$
more pronounced is the role of fluctuations (compare solid and dashed lines
which correspond to $M_{max}=12$ and $M_{max}=20$, respectively).
This observation is easy to explain:
The total number of operations in our algorithm is proportional to
$8M_{max}^2$ (for $M_{max}=N_{max}$) which means that the case $M_{max}=20$
requires almost three times more operations compared to the case
with $M_{max}=12$. The noise is accumulated as a function of elementary
operations, therefore to improve the fidelity of the preparation process
it is important to choose carefully $M_{max}$.
To be specific, for a given $\varepsilon$ we have to choose
the {\it minimal} values of $M_{max}$ and $N_{max}$ such that
\be
\sum_{m=0}^\infty
\sum_{n=0}^\infty |Q_{mn}|^2 -
\sum_{m=0}^{M_{max}}
\sum_{n=0}^{N_{max}} |Q_{mn}|^2 \leq \varepsilon.
\label{fin}
\ee

We can also use
Fig.~2 to illustrate the fact  that for a given value of $M_{max}$
the fidelity of the preparation may  depend on the target state.
To be specific, we have found that smaller the number
of nonzero amplitudes $Q_{mn}$ in the target state (\ref{st1}) the
higher the fidelity is for a given value of the range of
fluctuations $\delta$.
This behavior can be rather surprising as for synthesis of two states
``localized'' within the same region of the vibrational ``lattice'' $m,n$
we need comparable number of elementary operations.
Nevertheless, each nonzero $Q_{mn}$ after the synthesis is biased by
more or less the same error and, consequently,
the states with smaller number of nonzero amplitudes $Q_{mn}$ are less
sensitive to fluctuations. As already stated, this observation concerns
only states ``localized'' within the same region of the vibrational
lattice $m,n$ and with a comparable mean number of vibrational quanta.

\setcounter{figure}{1}
\begin{figure}
% Fig.2
\centerline{\epsfig{height=6.0cm,file=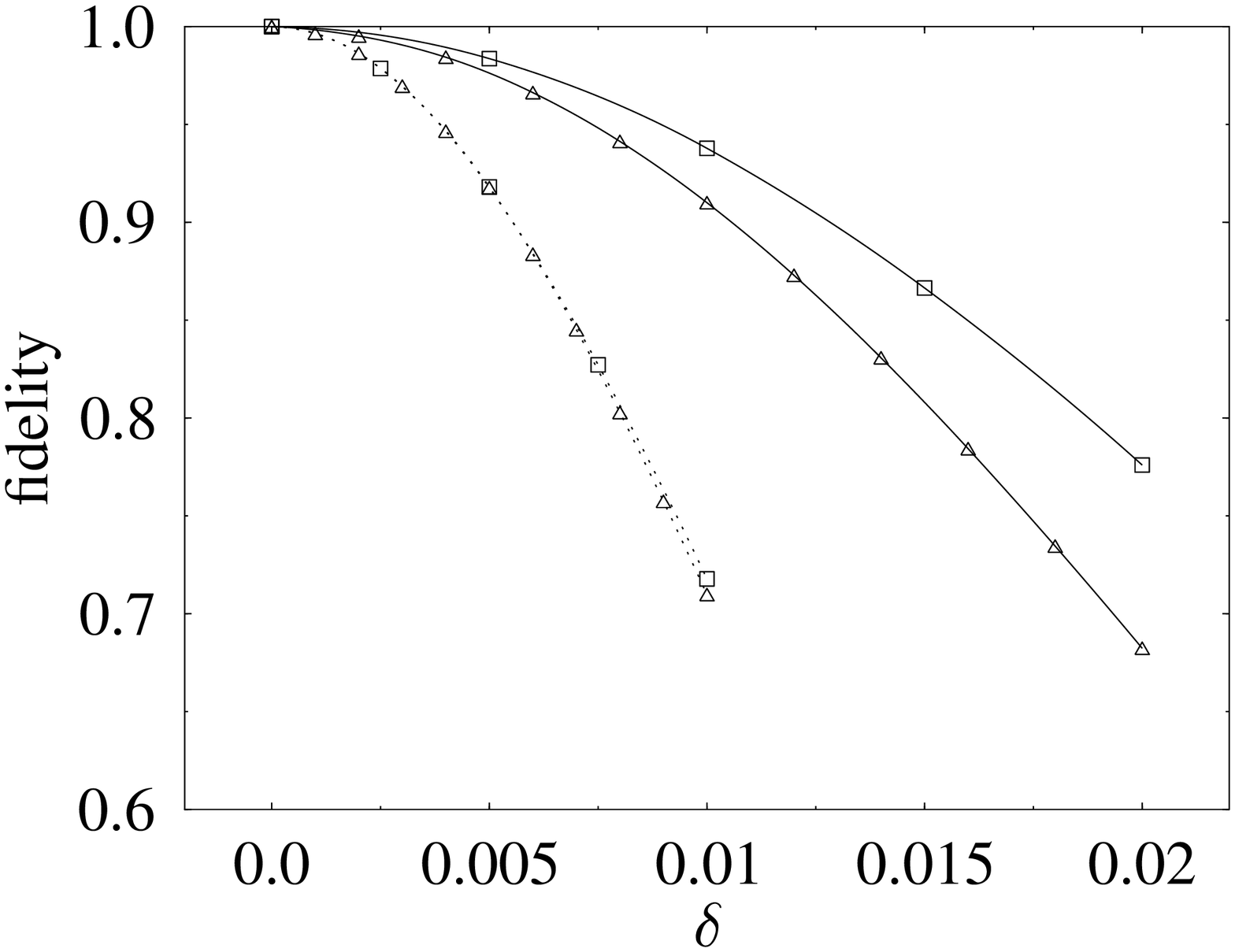}}
\caption{
\narrowtext
We plot fidelity given by Eq.(\protect\ref{fid})
as a function of the range of fluctuations $\delta$. We consider
two different target states,
two--mode cat--like state $|\Psi_{cat}\rangle$  (denoted by $\triangle$)
and two--mode correlated state $|\Psi_{corr}\rangle$
(denoted by $\fbox{\null}$) %%%$\square$
with $\alpha =2$.
We have considered two cases when  $M_{max}=12$ (solid lines)
and $M_{max}=20$ (dashed lines).
Comparing two solid lines we see that smaller is the number of nonzero
amplitudes $Q_{mn}$ in the target state (\ref{st1}) the larger is the
fidelity for a given value of $\delta$.
}
\end{figure}
%%%%%%%%%%%%%%%%%%%%%%%% FIGURE 2 %%%%%%%%%%%%%%%%%%%%%%%%%%%%%%%%%
%} %end_def

Finally, we briefly compare our algorithm with the one  proposed
by Gardiner, Cirac and Zoller \cite{Gardiner 1997}  in which
the number of operations in the preparation sequence is
growing {\it exponentially} as $2M_{max} \times 2^{M_{max}}$.
Therefore  fluctuations for $M_{max}$ large enough cause insurmountable
problem.
There is also another problem with this procedure. Namely, Gardiner, Cirac,
and Zoller have utilized only  two internal atomic levels
($|a\rangle$, $|b\rangle$) which results in the fact that their algorithm
is based on manipulations with vibrational states which are out of the
original Hilbert space specified by the cutoffs $M_{max}$
and $N_{max}$. In other words, manipulations with highly excited
vibrational states are required for construction of states with
relatively small number of vibrational quanta. This means that not
only the number of operations is exponentially growing but also
the number of vibrational quanta during the preparation procedure
may transiently exponentially increase.

We stress that our algorithm is associated with manipulations only within
the original subspace of the Hilbert space specified by $M_{max}$
and $N_{max}$. Moreover,
the number of operations grows only polynomially as $8M_{max}^2$.
The dimension of the two--mode Fock subspace from which the component
states are  does not increase during the preparation procedure.
This great reduction of number
of operations is due to the fact that we have employed
the third atomic level $|c\rangle$ in the preparation procedure.
Very recently a new 2-D preparation scheme was introduced by 
Kneer and Law \cite{Kneer 1998}.
The standing--wave laser field in $y$ direction induces a 
photon-number dependent interaction enabling thus to use 1-D schemes
selectively for particular subspaces with constant number of trap quanta 
in $y$ direction.  The number of operations scales also polynomially as 
$2M_{max}^2$ with only two electronic levels involved.

%%%%%%%%%%%%%%%%%%%%%%%% CONCLUSIONS %%%%%%%%%%%%%%%%%%%%%%%%%%%%%%%%
\section{Conclusion}
%%%%%%%%%%%%%%%%%%%%%%%%%%%%%%%%%%%%%%%%%%%%%%%%%%%%%%%%%%%%%%%%%%%%%

In this paper we have presented a universal algorithm for an efficient
deterministic preparation of an arbitrary two-mode bosonic state. We have
adopted this algorithm 
as a computer program \cite{www}. 
The proposed method can be generalized to 3-D trapping potential
and three--mode vibrational states. In this case one would need
four internal electronic levels and nine interaction channels
coupled to the 3-D vibrational field.
It can be shown that the number of operations required for quantum state
synthesis scales polynomially ($\sim M_{max}^3$).
Further generalization to multi-mode fields is possible \cite{Hladky}.

\acknowledgements

We thank Jason Twamley for helpful discussions.

%%%%%%%%%%%%%%%%%%%%%%% REFERENCES %%%%%%%%%%%%%%%%%%%%%%%%%%%%%%%
%\vspace{-0.5cm}

\end{multicols}

\end{document}